\documentclass{aa}
\usepackage{graphicx}
\usepackage{natbib}
\bibpunct{(}{)}{;}{a}{}{,}

\newcommand{\aap}{A\&A}
\newcommand{\apj}{ApJ}
\newcommand{\mnras}{MNRAS}

\begin{document}

\title{Tree-Structured Grid Model of Line and Polarization Variability from Massive
Binaries}

\author{R. Kurosawa \and D. J. Hillier}

\institute{Department of Physics and Astronomy, University of Pittsburgh, 3941 O'Hara
Street, Pittsburgh, PA 15260}

\offprints{R. Kurosawa}

\mail{kurosawa@phyast.pitt.edu}

\date{Received  DATE  / Accepted DATE }

\abstract{
We have developed a 3-D Monte Carlo radiative transfer model which computes
line and continuum polarization variability for a binary system with an optically
thick non-axisymmetric envelope. This allows us to investigate the complex (phase-locked)
line and continuum polarization variability features displayed by many massive
binaries: W-R+O, O+O, etc. An 8-way tree data structure constructed via
a {}``cell-splitting{}'' method allows for high precision with efficient use of
computer resources. The model is not restricted to binary systems; it can easily
be adapted to a system with an arbitrary density distribution and large density
gradients. 
As an application to a real system, the phase dependent Stokes 
parameters \( \left( I,\, Q,\, U\right)  \) and the phase dependent \ion{He}{i} (\(
\lambda  \)5876) profiles of the massive binary
system \object{V444~Cyg} (WN5+O6 III-V) are computed.
}

\titlerunning{Tree-Structured Grid Polarization Model}
\maketitle

\section{Introduction}

\label{sec:Introduction}

Linear polarization of light is often used to study the geometrical distribution
of gaseous/dusty materials around various types of astronomical objects, e.g.,
close binaries, Be stars, Wolf-Rayet (W-R) stars, Luminous Blue Variables (LBVs),
supernovae, Seyfert galaxies, quasars (QSOs), and Young Stellar Objects (YSOs).
When the angular size of an object is too small to be resolved by ground-based
or space telescopes, spectropolarimetry and narrow band polarimetry provide
useful information on the geometry of circumstellar matter. 

Several analytic models use simplifying assumptions to predict the continuum
polarization levels arising from electron scattering envelopes. For example,
\citet{brown:1977} derived an expression for the degree of polarization produced
in an optically thin axisymmetric envelope with an embedded point source. \citet{brown:1978}
extended this model, deriving the expressions for the polarization arising from
a binary system embedded in an optically thin envelope. Adopting this model,
\citet{brown:1989} considered the effect of the finite size of the light source.
Unfortunately, these 2-D models lack an ability to investigate more complex
systems such as close binaries with colliding stellar winds, molecular clouds
with fractal structure, and the tails of non-circular accretion flows. Further,
they can not deal with the realistic inhomogeneities present in many scattering
envelopes, and they are not suitable for optically thick atmospheres with multiple
scattering. 

In order to investigate such complicated systems, we abandon the simplifying
assumptions and use a full three dimensional model that can handle an optically
thick atmosphere. Formulating and solving the radiative transfer equation for
an arbitrary geometry is difficult even for continuum radiation. The presence
of additional scattering integrals in the expression for the source function
and the need to solve several coupled transfer equations for the Stokes parameters
tremendously complicates the problem. The flexibility of Monte Carlo methods
of radiative transfer are particularly useful for such problems despite the
fact that they may not be the most efficient solution method. For example, asymmetries
in the radiation field, non-spherical geometry, and a complex distribution of
the clumps can be readily incorporated. The issue of efficiency is less severe
due to the significant advancement of computer technology. In the near future,
it may be possible to perform non-LTE radiative transfer calculations in 3-D
including thousands of lines from multi-atomic species utilizing some of the
ideas described by \citet{bernes:1979} and \citet{li:1996}. Recently, \citet{lucy:1999}
developed an improved Monte Carlo technique for non-LTE radiative transfer calculations
applied to a synthetic supernova spectra model.

Our main reason for developing a 3-D model is to investigate the colliding wind
(CW) interaction zone. Such zones are relatively common among massive binaries
(W-R+O, O+O binaries). \citet{koenigsberger:1985} found, using low resolution
IUE spectra, that 3 out of 6 selected W-R+O binaries showed evidence for CWs.
\citet{shore:1988}, \citet{marchenko:1994} and \citet{marchenko:1997} presented
detailed studies of CW related variability seen in high-resolution UV and optical
data of \object{V444~Cyg} (WN5+O6 III-V). \citet{stevens:1999} presented observations
of the phase dependent \ion{He}{i} (1.0830-\( \mu  \)m) line profile in six
different W-R binary systems including \object{V444~Cyg}. Using a simple model,
they explained the observed variability in terms of the wind-wind interaction.
Further theoretical aspects of the relationship between X-ray variability and
CWs have been developed by \citet{luo:1990} and \citet{stevens:1992}. More recently,
detailed hydrodynamic models of colliding winds have be used by \citet{stevens:1994},
\citet{gayley:1997} and \citet{pittard:1998}. Simulations of the colliding winds
in \object{V444~Cyg} are given in \citet{pittard:1999}.

Examples of 3-D Monte Carlo radiative transfer models recently developed are
\citet{witt:1996}, \citet{pagani:1998}, \citet{stevens:1999}, \citet{wolf:1999}
and \citet{harries:2000}. \citet{juvela:1997} presented their 1-3 dimensional
Monte Carlo models of emission from clumpy molecular clouds. \citet{wolf:1999} developed
a \emph{self-consistent} 3-D continuum Monte Carlo model in which the dust temperature
and the radiation field are iterated to consistency. These 3-D models, with
the exceptions of \citet{wolf:1999} and \citet{harries:2000}, use regular {}``cubic{}''
grids and are not suitable for a model with a large density gradient or a large
dynamic range of density. These would require much finer grid sizes (and hence
more memory) to resolve the smaller structures. \citet{wolf:1999} and \citet{harries:2000}
used 3-D grids in spherical coordinates. Their grids are evenly spaced in polar
and azimuthal angles, but logarithmically spaced in the radial direction. This
type of grid is useful for a system that does not deviate much from spherical
or axisymmetric symmetry, but it would not be adequate for a binary system with
colliding stellar winds or for a binary system with accretion flow from a companion.

Starting from the 2-D Monte Carlo model of \citet{hillier:1991}, we have developed
a 3-D Monte Carlo code to calculate the continuum and line polarizations produced
by the scattering of light in an arbitrary geometry. The model can predict the
variability features associated with the orbital motion of a binary system,
such as the polarization level, flux level, and line profile shapes. The model
can treat a finite size stellar disk, multiple scattering, absorption of continuum
photons by a line, and emission from multiple light sources (extended or not).
Higher precision is achieved with fewer grid points by using a {}``cell-splitting{}''
method whose application to this type of problem was discussed by \citet{wolf:1999}.
We extended the idea of {}``cell-splitting{}'' to a simple 8-way tree data
structure to construct the model grids. With this data structure, we can quickly
access the data stored at any grid point, leading to a faster, more efficient,
numerical code. This is essential to our model because of the high dimensionality
and because no assumptions are made regarding the symmetry of the model. A logarithmic
grid, like those commonly used in spherical and axisymmetric codes, is not readily
implementable.

In \S~\ref{sec:Model}, we describe the details of our model. Various accuracy
and efficiency tests are given in \S~\ref{sec:Tests}. The model is applied
to the massive binary \object{V444~Cyg} (WN5+O6) as an example calculation
for a real system in \S~\ref{sec:V444Example}. The conclusions are given in
\S~\ref{sec:Conclusion}.

\section{Model}

\label{sec:Model}

\subsection{Polarization}

\label{subsec:Polarization}

A general method of polarization calculation via the Monte Carlo method is described
in \citet{modali:1972}, \citet{sandford:1973}, \citet{warrensmith:1983} and
\citet{hillier:1991}. An initial photon beam emitted from a light source is
usually assumed to be unpolarized, with the Stokes parameters given by \( (I,Q,U,V)=(I_{o},0,0,0) \).
After each scattering, the Stokes parameters change according to the Mueller
matrix, \( M(\theta ,\phi ) \) where \( \theta  \) and \( \phi  \) are the
polar and azimuthal scattering angles respectively.

\[
\left( \begin{array}{c}
I'\\
Q'\\
U'\\
V'
\end{array}\right) \propto M(\theta ,\phi )\left( \begin{array}{c}
I\\
Q\\
U\\
V
\end{array}\right) \]
 In the case of Thomson scattering , \( M \) has no \( \phi  \) dependency,
and it becomes 

\[
M=\left( \begin{array}{cccc}
\frac{1}{2}(1+\mu ^{2}) & -\frac{1}{2}(1-\mu ^{2}) & 0 & 0\\
-\frac{1}{2}(1-\mu ^{2}) & \frac{1}{2}(1+\mu ^{2}) & 0 & 0\\
0 & 0 & \mu  & 0\\
0 & 0 & 0 & \mu 
\end{array}\right) \]
where \( \mu =\cos \theta  \). 

A coordinate transformation of the Stokes parameter vector \( \Lambda ^{T}=\left( I,Q,U,V\right)  \)
to a fixed reference frame is necessary after each scattering. \citet{chandrasekhar:1960}
describes how this transformation is done. After several scattering events,
the photon beam will reach the model boundary (cubic in our case), and the \( \Lambda ^{T} \)
vector will then be projected onto the plane of an observer. The
reader is refer to \citet{hillier:1991}
for details.

\subsection{Coordinates }

\label{subsec:Coordinates}

\begin{figure}
{\par\centering \resizebox*{!}{0.3\textheight}{\includegraphics{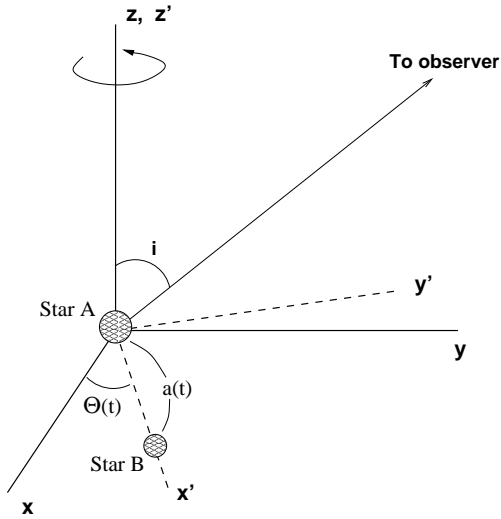}} \par}

\caption{\label{fig:coordinates} This illustrates the two coordinate systems used in
the models. The binary coordinates (x', y' z') rotates around the primary star
coordinates (x, y, z) through the common z or z' axis. A rotation angle (\protect\( \Theta \protect \))
and a binary separation (\protect\( a\protect \)) are functions of orbital
phase \protect\( t\protect \). For a circular orbit, they are simply, \protect\( \Theta \left( t\right) =2\pi \protect \)
and \protect\( a=\protect \)constant. An observer is placed on the y-z plane
with the viewing angle \protect\( i\protect \). The primary star (Star A) is
placed at the origin of the two coordinate systems, and the secondary star (Star
B) is placed on the x' axis. With this configuration, the orbit of the binary
is confined on x-y plane.}
\end{figure}

For the model discussed here, we assume that the density distribution is \emph{co-rotating}
with the binary system, and the orbit is \emph{circular}. Two coordinate systems
are used in the model (see Fig.~\ref{fig:coordinates}). One is the primary
star coordinates (x, y, z), and the other is the binary coordinates (x', y',
z'). The binary coordinate system is a rotating coordinate system, and it is
chosen such that the stars and the density are fixed (or independent of orbital
phases) in that coordinate system. The orbital plane is assigned to be on the
x-y plane of the primary star coordinates. The primary star (Star A) is placed
at the origin of the two coordinate systems, and the secondary star (Star B)
is placed on the x' axis. The relative position of two coordinate systems changes
as a function of orbital phase (\( t \)). When \( t=0 \), the two
coordinate systems coincide. For \( t>0 \), the primed (binary)
coordinate system rotates around 
the z axis of the unprimed (primary star) coordinates making the angle between
x \& x' and y \& y' to be \( \Theta \left( t\right)  \). For a circular orbit
the relative angle is simply, \( \Theta \left( t\right) =2\pi \, t \). In general,
the separation distance \( a \) is a function of \( t \), but for a circular
orbit \( a \) remains constant. By our convention, an observer is always on
the y-z plane with an inclination angle \( i \) measured from z or z' axis.

In the primary star (unprimed) coordinates, the locations of the stars, as a
function of \( t \), become 

\begin{equation}
\label{eq:PosStarA}
\left( \begin{array}{c}
x_{A}(t)\\
y_{A}(t)\\
z_{A}(t)
\end{array}\right) =\left( \begin{array}{c}
0\\
0\\
0
\end{array}\right) 
\end{equation}
\label{eq:PositionStarA}

and

\begin{equation}
\label{eq:PosStarB}
\left( \begin{array}{c}
x_{B}(t)\\
y_{B}(t)\\
z_{B}(t)
\end{array}\right) =\left( \begin{array}{c}
a\cos \left\{ 2\pi \left( t+t_{0}\right) \right\} \\
a\sin \left\{ 2\pi \left( t+t_{0}\right) \right\} \\
0
\end{array}\right) 
\end{equation}
\label{eq:PositionStarB} where \( 2\pi t \) is the relative angle between
the two coordinates and \( t_{0} \) is an initial phase offset. In our model
\( t_{0}=-1/4 \) is used so that when \( t=0 \), the secondary star (Star
B) is behind the primary star (Star A) from an observer's perspective. Eqs.~\ref{eq:PosStarA}
and \ref{eq:PosStarB} can be easily modified to an orbit with non-zero eccentricity
for a general binary orbit.

\subsection{Grid Selection}

\label{subsec:GridSelection}

When the gradient in the opacity field is very large, a logarithmic scale in
the radial direction can be used to increase computational accuracy of the optical
depth if the system is spherically or axially symmetric. For the case of an
arbitrary geometry, there is no simple way to construct a logarithmic scale. However,
an efficient cubic grid (consisting of cubic \emph{cells}) can be constructed by subdividing a
cube into smaller cubes where the value of the opacity/emissivity is larger
than a given threshold value. \citet{wolf:1999} introduced the basic idea for
this type of grid method to the radiative transfer problem. We do not know the
exact algorithm they used, but we have utilized the 8-way tree data structure
to construct the grids and to store the values of the opacities. 
In addition to being conceptually simple, the advantages of using a
tree data structure include increases in optical depth calculation
accuracy, computer memory efficiency, and data access time.  In the
following subsections, we describe the algorithm for constructing the
tree-structured grid and illustrate its efficiency.

\subsection{Tree Data Structure Construction Algorithm}

\begin{figure}
{\par\centering \resizebox*{!}{0.3\textheight}{\includegraphics{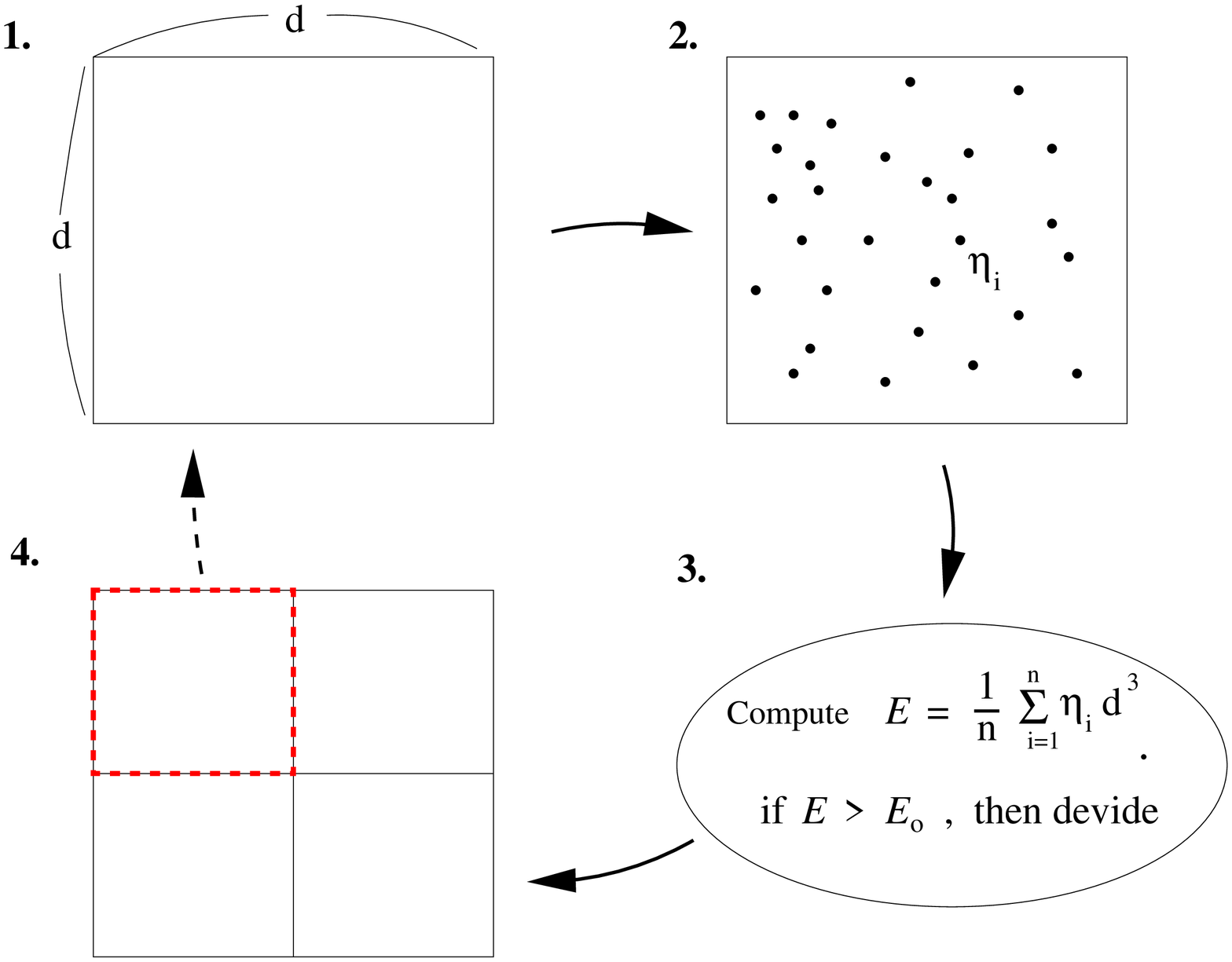}} \par}

\caption{\label{fig:grids} This diagram shows how a cell in a model space will be split
into 8 subcells. 1.\,Consider a cubic cell with length {}``d.{}'' 2.\,Choose
n random locations in the cell, and evaluate the emissivity value 
\protect\( (\eta_{i})\protect \)
at each location. Average these emissivities to obtain 
\protect\( E \protect \). 3.\,If \protect\( E\protect \) is greater
than a threshold value \protect\( E_\mathrm{o}\protect \), then 4.\,Divide the
cell into 8 equal size cubic cells. The program will repeat this procedure recursively
until all the cells satisfy the condition: \protect\( E < E_\mathrm{o}\protect \).
Note that the figure shows the cell split into only 4 cells because the 3-D
cell is projected onto a 2-D plane for clarity.}
\end{figure}

The data structure used here is similar to a {}``binary tree{}'' in which
one node splits into two children nodes, but ours splits into eight. The steps
for the tree construction are summarized below. Fig.~\ref{fig:grids} gives the
flow chart of the steps.

\begin{enumerate}
\item Start from a cubic box (\emph{root node}) which holds the whole density structure.
\item Pick (100-1000) random locations, \( \mathbf{r}=(x,y,z \)), in the cell, and
compute 
\begin{equation}
\label{eq:CellValue}
E =\frac{1}{n}\sum ^{n}_{i=1}\eta _{i}^{p}\, d^{3}
\end{equation}
 where \( \eta _{i} \) is the emissivity at \( i \)-th random location, \( n \)
is the total number of the random locations, \( d \) is the size of the cell
and \( p \) is the index of scalability. 
\item If \( E \) is greater than \( E_{\mathrm{o}} \) which is a user
specified parameter, divide the box into 8 equal-size cubes (\emph{child nodes}).
If \( E \) is smaller than \( E_{\mathrm{o}} \), do nothing.
\item For each cell just created, repeat Steps 2 and 3 until all cells
have \( E < E_{\mathrm{o}} \).
--- The lowest level cells are often called {}``\emph{leaf(s)}{}'' (\emph{leaves}).
\end{enumerate}
The index \( p \) in Eq.~\ref{eq:CellValue} controls how fast the cell should
be split with \( \eta  \), and normally \( p=1 \) is used. 
When $p=1$, $E$ represents the average {}``emission measure{}'' of the
cell since it simply becomes the product of the average emissivity and the
volume of the cell.
The emissivity at a given point can be evaluated from one or a
combination of the following: 1.\,an analytical formula, 2.\,an
interpolation of the output from a radiative transfer model and
3.\,an interpolation of the output from a hydrodynamical model. 
Other more complex algorithms could be used to optimize the
computation for a specific problem. 

\begin{figure}
{\par\centering \resizebox*{!}{0.3\textheight}{\includegraphics{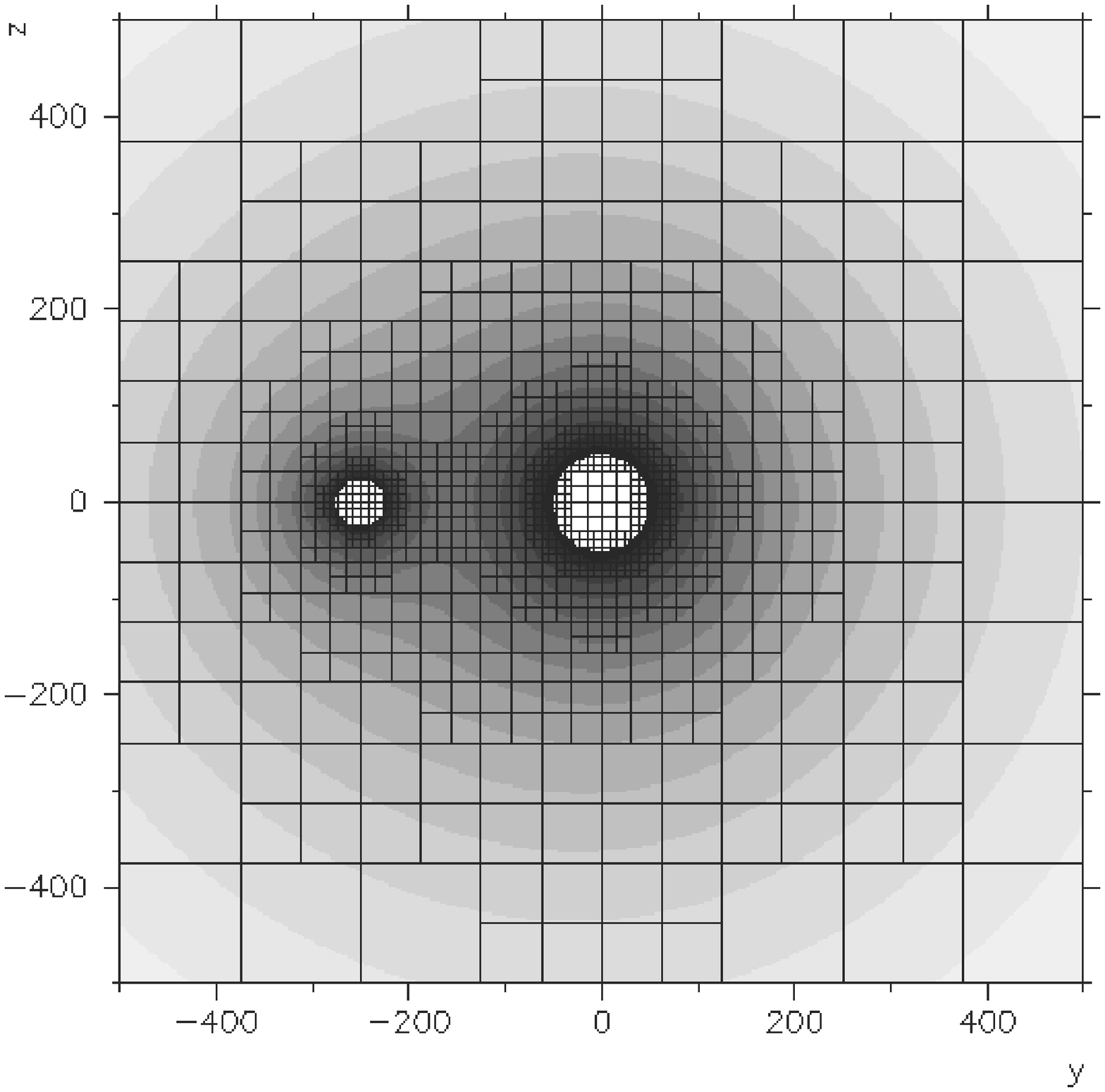}} \par}

{\par\centering \resizebox*{!}{0.3\textheight}{\includegraphics{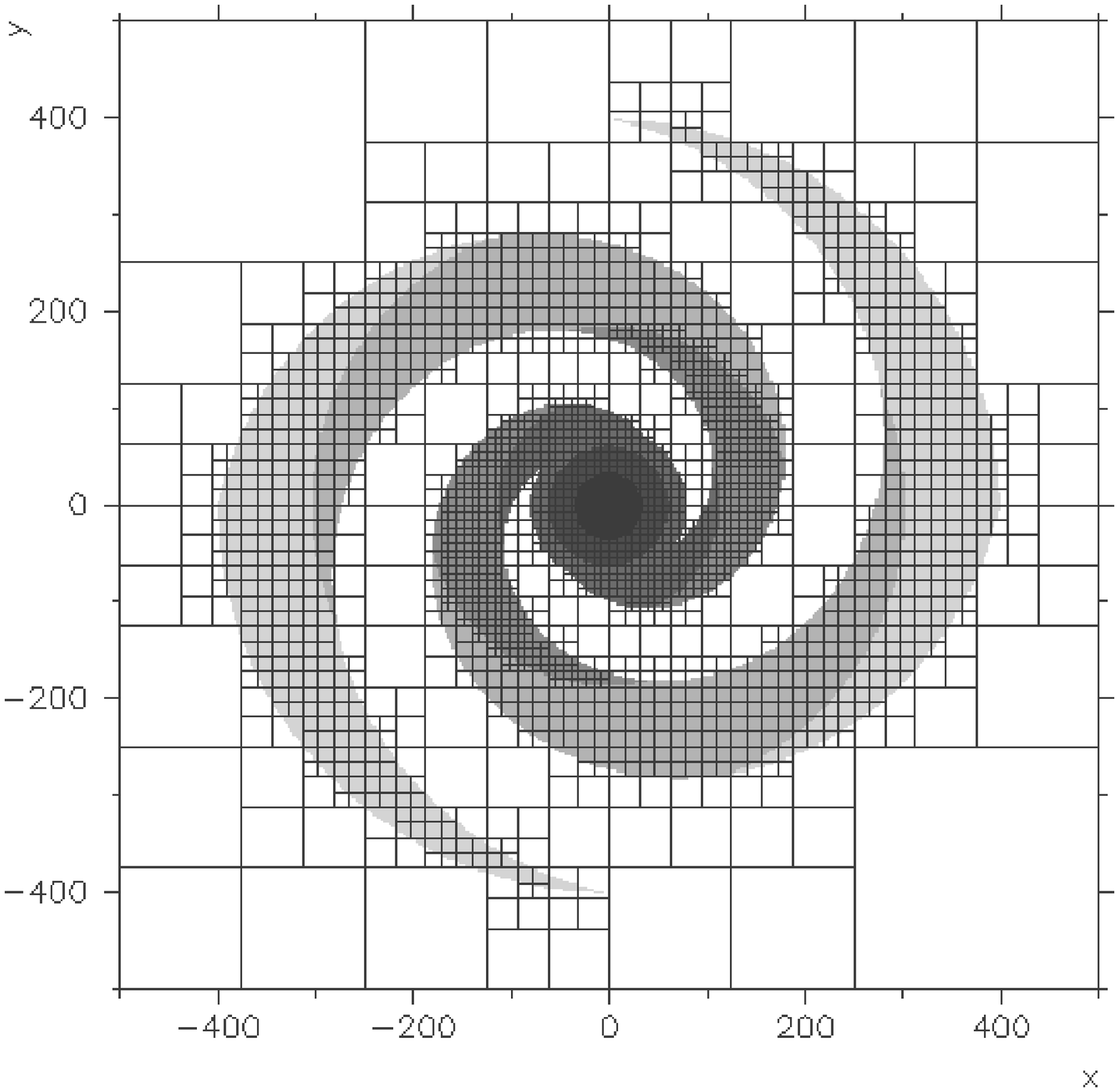}} \par}

\caption{\label{fig:demo01} Examples of the tree structured grids. Top: a binary system
with different stellar radii surrounded by a spherical atmosphere around each
star. Bottom: a two-arm pseudo spiral galaxy. The density decreases as \protect\( \propto 1/r^{2}\protect \)
for \protect\( r>R_{b}\protect \) and it is constant for a \protect\( r<R_{b}\protect \)
where \protect\( R_{b}\protect \) is the bulge radius. Note that the real density
distributions are 3-D, but here they are depicted in 2-D for simplicity. In
both cases, the grid size naturally decreases for the high density regions.}
\end{figure}

Once the data structure is built, the access/search for a given cell can be
done recursively, resulting a faster code. Examples of the resulting
grids are shown in Fig.~\ref{fig:demo01}. The top diagram
in Fig.~\ref{fig:demo01} shows the grids constructed for two stars with different
sizes having spherically distributed gas around each star. The figure clearly
shows that the grid size becomes naturally smaller where the density is higher.
The grids used for this density is in 3-D, but the diagram shows only the cross
section of the y-z plane for clarity. In the bottom diagram of Fig.~\ref{fig:demo01},
the grids are overlaid with a pseudo-spiral galaxy density distribution (3-D
but drawn in 2-D). The density of the disk is inversely proportional to the
distance from the center of the galaxy except for the bulge region where the
density is constant. The grids nicely follow the spiral arms, and become smaller
as they approach to the center.

\subsection{Searching for a Cell}
\label{subsec:SearchTree}

\begin{figure}
{\par\centering \resizebox*{!}{0.45\textheight}{\includegraphics{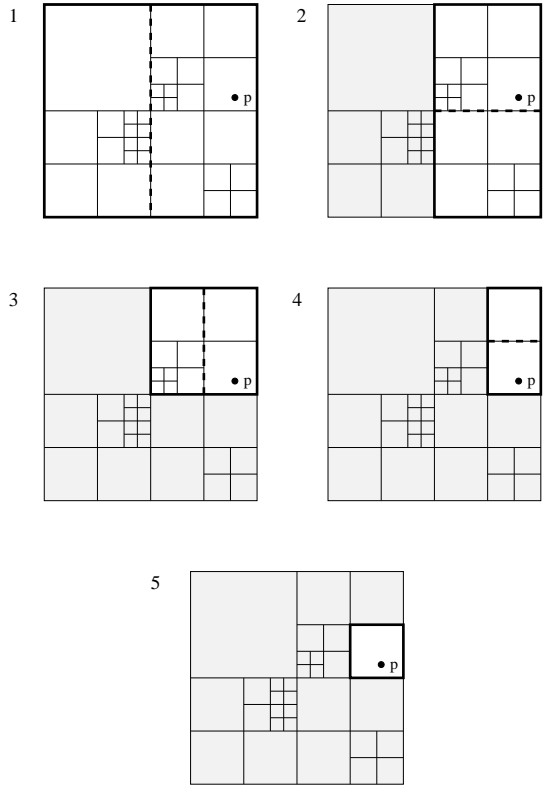}} \par}

\caption{\label{fig:search_tree} An illustration of how the cell which
contains a point ($p$) in the model space can be found in the tree
data structure.  The figure demonstrates the searching steps for the 2-D case
for clarity, but the same method can be applied to the 3-D case.
The thick solid line indicates the cell that is focused in each step.  
The thick dotted lines represent the border used 
to check which side of the cell point p occupies. 
The corresponding instructions for each step are summarized in
\S~\ref{subsec:SearchTree}.}
\end{figure}

An example of how to find the cell which contains a point ($p$) 
in the tree data structure is illustrated in
Fig.~\ref{fig:search_tree}. The steps shown here are for the 2-D case
for clarity, and the same method can be applied to the 3-D case.
The figure shows the following steps: 

\begin{enumerate}
\item   Starting from the root cell (the outer most cell), outlined
	with a thick solid line, 
	check whether the point is in the left half or the right half
 	of the cell. 
\item   The point is in the right half of the cell; hence, dismiss all
	the subcells in the left half.  
\item	Further, check whether 
	the point is in the upper half or the lower half of the cell
	outlined with a thick solid line.	
   	The point is in the upper half of the cell; hence,  dismiss
	all the subcells in the lower half.  
\item	Further, check whether the point is in the left half or
	the right half of the cell outlined with a thick solid line.
\item   The point is in the right half of the cell; hence, dismiss all
	the subcells in the left half.
	Further, check whether the point is in the upper or
	the lower of the cell outlined with a thick solid line.
\item  The point is in the lower half of the cell. 
	Since this is the lowest level
	(leaf node) of the tree, the search stops here. 
       The leaf cell containing point $p$ is found.
\end{enumerate}

In our algorithm, whether a point belongs to the left half or 
the right half of the cell and whether a point belongs to the upper
half or the lower half of the cell are checked simultaneously. Therefore, steps 1
and 2 are considered to be just one step, as are steps 3 and 4. 
In other words, the total number of steps required to find the
cell including point $p$ is considered to be ``two'' in this example.

\subsection{Performance Check}

\begin{figure}
{\par\centering \resizebox*{!}{0.3\textheight}{\includegraphics{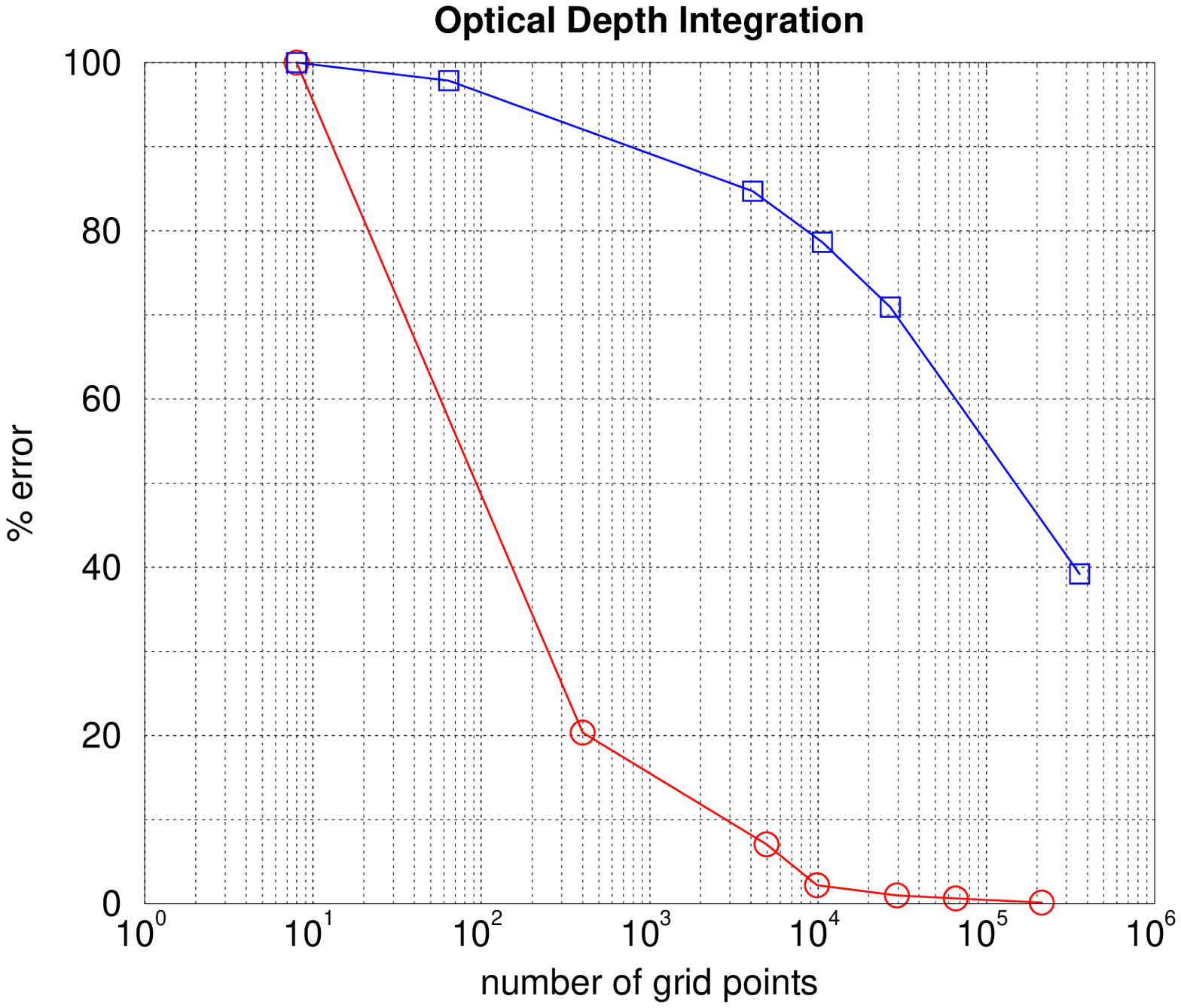}} \par}

{\par\centering \resizebox*{!}{0.3\textheight}{\includegraphics{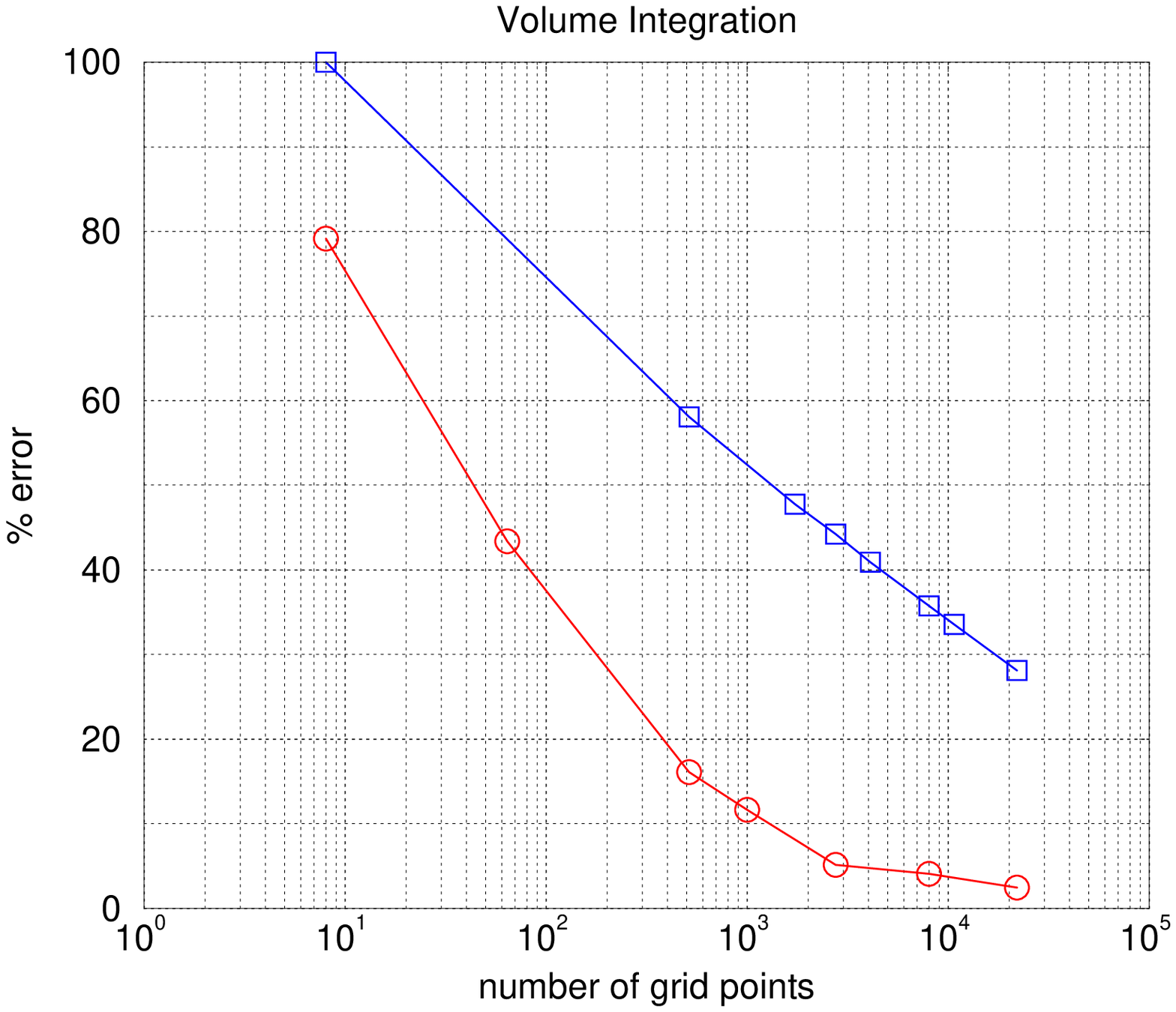}} \par}

\caption{\label{fig:accuracy} Top: Percentage errors of optical depth integrals using
a regular grid (squares) and using a tree structured grid (circles) v.s. 
total number of grid points. 
The opacity is set to be \protect\( \chi =\chi _{0}/r^{4}\protect \), and
the optical depth is calculated from the surface of a star (\protect\( r=1R_{*}\protect \))
to the outer boundary of the model (\protect\( r=100R_{*}\protect \)). Bottom:
Percentage errors of volume integrals \protect\( \left( \sum _{all\, cells}\rho _{i}dV_{i}\right) \protect \)
using a regular grid (squares) and using a tree structured grid
(circles) v.s. total number of grid
points. The density is assumed to be \protect\( \rho =\rho _{o}/r^{2}\protect \),
the integration limits are \protect\( r_{\mathrm{min}}=1R_{*}\protect \) and
\protect\( r_{\mathrm{max}}=100R_{*}\protect \) where \protect\( R_{*}\protect \)
is the radius of a star.}
\end{figure}

To demonstrate the efficiency of the tree structure,\footnote{%
For this simple density distribution, the cylindrical coordinates \( \left( r,\, \phi ,\, z\right)  \)
with logarithmic spacing in \( r \) direction would be very efficient, but
such a grid scheme is too restricted for an arbitrary density distribution.
} the optical depth and the volume integral of density were computed with a (single-size)
regular cubic grid and a tree structure grid separately. The functional forms
of the opacity and the density used are \( \chi \left( r\right) =\chi _{0}/r^{4} \)
and \( \rho \left( r\right) =\rho _{0}/r^{2} \) respectively. The top graph
in Fig.~\ref{fig:accuracy} shows the percentage error of the optical depth
calculations with the tree structure grid and that with one-size cubic grids
as a function of {}``total{}'' number of grid points. Similarly, the bottom
graph in Fig.~\ref{fig:accuracy} shows the percentage error of volume integrals
as a function of total number of grid points. Note that no special integration
weights were used in both cases.

The figure shows the error decreases very quickly as the total number of grid/node
points increases for the tree structure grids. As a consequence, much smaller
number of points is required for the tree method to achieve the same order
of accuracy, and thus the method requires much less computer memory. This enables
us to do a realistic 3-D simulation even on a regular PC with a reasonable amount
of RAM. 

The number of steps in a grid searching routine depends on the complexities
of the data structure, but in the case of the grids constructed with the inverse-square
density function with the total grid points of \( \sim 10^{6} \), the average
number of searching steps is about 5. Although not implemented to the current
model, the searching could be made more efficient by searching upward from the
current position of a photon rather than searching downward from the root node
each time since the cells along a photon trajectory are correlated.

\subsection{Optical Depth}

\label{subsec:OpticalDepth}

\begin{figure}
{\par\centering \resizebox*{!}{0.3\textheight}{\includegraphics{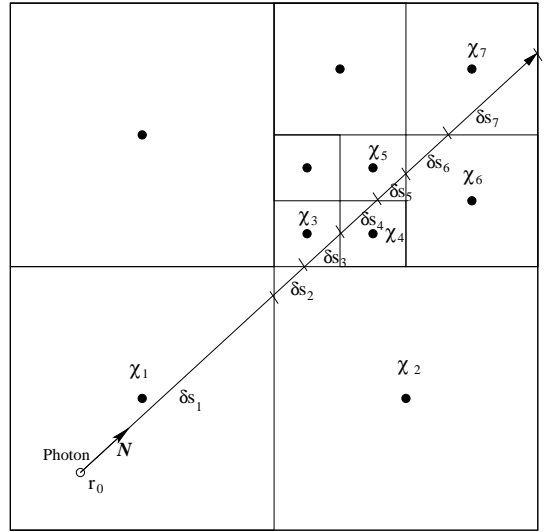}} \par}

\caption{\label{fig:tau01} This diagram shows a photon at \protect\( \mathbf{r}_{0}\protect \)
moving in direction \protect\( \mathbf{N}\protect \) within the model space.
The beam of the photon intersects with different sizes of cells before it reaches
the outer boundary. The value of the opacity of each cell is assigned at the
center of the cell, and the optical depth is calculated simply as the sum of
the product of the cell value (\protect\( \chi _{i}\protect \)) and the line
segment length (\protect\( \delta s_{i}\protect \)) as in Eq.~\ref{eq:tau}. }
\end{figure}

Suppose a photon at \( \mathbf{r}_{o}=(x_{o,}y_{o},z_{o}) \) is moving in the
direction of \( \mathbf{N} \) as shown in Fig.~\ref{fig:tau01}. This
photon will 
most likely intersect with many cells (or grids) before it reaches the outer
boundary. Calculation of the optical depth for a photon which is at \textbf{\( \mathbf{r}_{o} \)}
and moving in direction \( \mathbf{N} \) is done by replacing the integral
along the ray with the sum of the product of the cell opacity (\( \chi _{i} \))
with the line length (\( \delta s_{i} \)) of the photon beam in the
cell, yielding
\begin{equation}
\label{eq:tau}
\tau =\int \chi (s)\, ds \approx \sum ^{m}_{i=1}\chi _{i}\, \delta s_{i}
\end{equation}
where \( \chi _{i} \) is the sum of thermal and electron scattering opacities
for a continuum photon: 
\[
\chi _{i}=\chi _{i}^{th}+\chi _{i}^{es}
\]
and \( m \) is the number of cells which intersect with the photon. In order
to perform the summation in Eq.~\ref{eq:tau}, we first need to find with which
cells the photon will intersect. Then, the opacity values stored in these cells
must be extracted. The location of the intersections on the cells must be found
in order to calculate \( \delta s_{i} \) in Eq.~\ref{eq:tau}. As mentioned
before, the searching for the intersecting cells and the values stored in the
cells are relatively fast since the data is stored in the tree data structure.

Finding the intersections of a photon beam with model grids is relatively simple
for a spherically symmetric system. In the tree grid case, the following must
be done to find the intersecting points:

\begin{enumerate}
\item For all the planes of the cells which could have intersected with this photon,
the following set of equations must be solved for \( (x,y,z) \). The first one
is the equation of the line along which the photon would move, and the second
is the equation of the plane.\begin{equation}
\label{eq:line_xyz}
\frac{(x-x_{0})}{N_{x}}=\frac{(y-y_{0})}{N_{y}}=\frac{(z-z_{0})}{N_{z}}
\end{equation}
\begin{equation}
\label{eq:plane+xyz}
(x-x_{c})\, n_{x}+(y-y_{c})\, n_{y}+(z-z_{c})\, n_{z}=0
\end{equation}
where \( (x_{c},y_{c},z_{c}) \) is a point on a face of the cube, \( (x_{0},y_{0},z_{0}) \)
is the current location of photon, \( \mathbf{n}=(n_{x},n_{y},n_{z}) \) is
the normal vector of a face, and \( \mathbf{N}=(N_{x},N_{y},N_{z}) \) is the
directional vector of the photon.
\item There may be up to three such intersections (since the ranges of the coordinates
are not restricted in Eqs.~\ref{eq:line_xyz} and \ref{eq:plane+xyz}), but
we are only interested in the first intersection which a photon encounters.
The distance to each intersection must be calculated, and the smallest
distance is used to compute \( \delta s_{i} \) in
Eq.~\ref{eq:tau}. 
\end{enumerate}

\begin{figure}

{\par\centering \resizebox*{!}{0.3\textheight}{\includegraphics{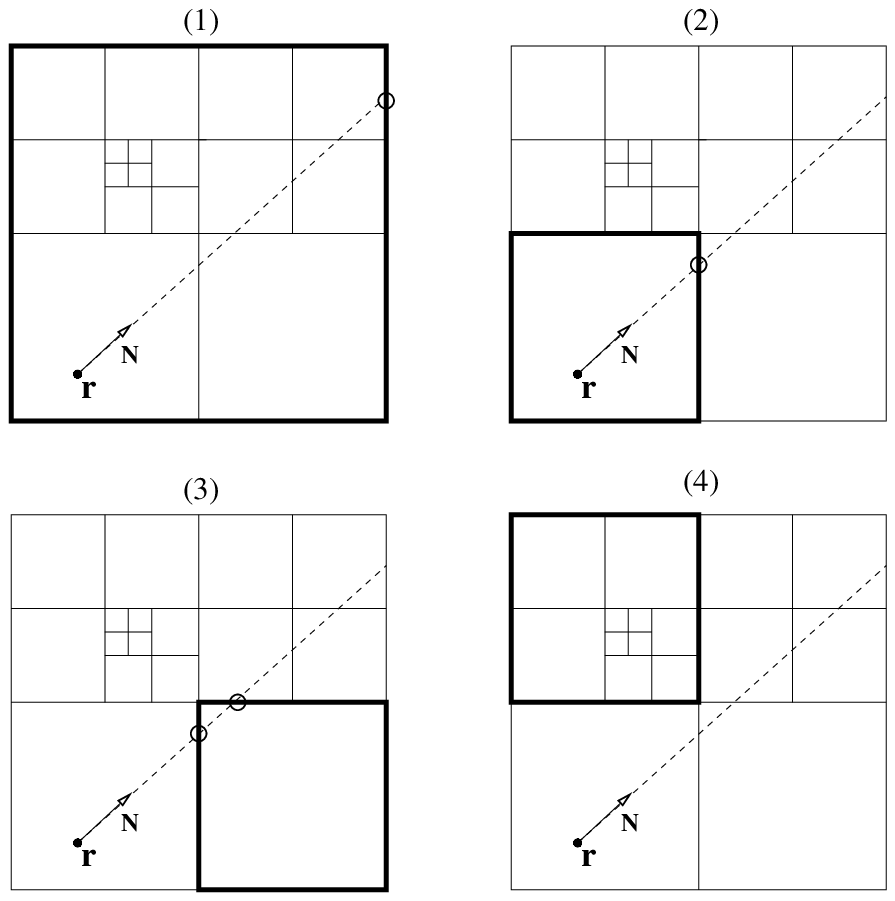}} \par}
{\par\centering \resizebox*{!}{0.45\textheight}{\includegraphics{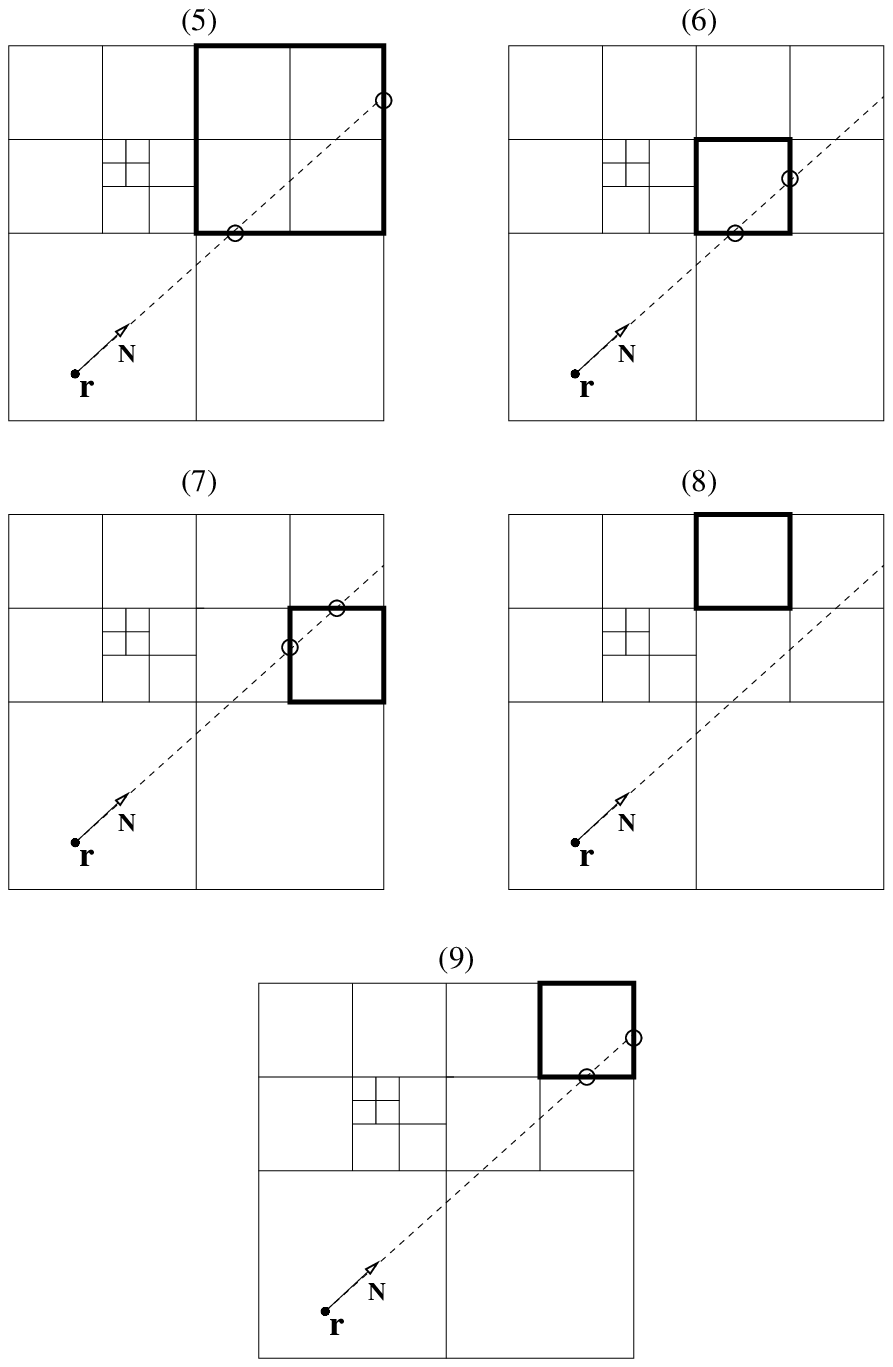}} \par}

\caption{\label{fig:optical_depth} An illustration of how the optical
depth between the outer boundary (the largest cell) and 
the current location ($\mathbf{r}$) of the photon 
moving in the direction $\mathbf{N}$ is calculated. 
The corresponding instructions for each step are
summarized in \S~\ref{subsec:OpticalDepth}.  The thick
solid lines indicate the important cell in each step. The
filled circle is the current location the photon, and the open circles are
the intersections of the photon beam and the cell of current
interest.} 
\end{figure}

Fig.~\ref{fig:optical_depth} shows a simple example of how to calculate 
the optical depth between the current photon location ($\mathbf{r}$) 
and the outer boundary (the largest cell in the figure). 
The figure is again depicted in 2-D for clarity, but the same
method can be used for the 3-D case.  The thick solid line indicates
the cell of current interest.  The corresponding steps in the
figure are the following:

\begin{enumerate}
\item 	There is only one intersection of the ray and the current cell 
	(thick line). Since this is not a leaf cell (the lowest level
	cell), descend to the children cells.
\item 	There is one intersection, and this is a leaf cell. 
	Find the distance ($\delta s_{1}$) between
	the photon position ($\mathbf{r}$) and the intersection. Compute \( d\tau
	_{1}=\chi _{1}\delta s_{1} \). Save $d\tau_{1}$.
\item 	There are two intersections, and this is a leaf cell. 
	Find the distance ($\delta s_{2}$) between
	the two intersections. Compute \( d\tau _{2}=\chi _{2}\delta
	s_{2} \).  Save $d\tau_{2}$.
\item 	There is no intersection, and this is not a leaf cell.
	Dismiss all the subcells.  
\item 	There are two intersections, but this is not a leaf cell. 
	Descend to the children nodes.
\item 	There are two intersections, and this is a leaf cell. 
	Find the distance ($\delta s_{3}$) between the two intersections. 
	Compute \( d\tau _{3}=\chi _{3}\delta s_{3} \). 
	Save \( d\tau _{3} \).
\item 	There are two intersections, and 
	this is a leaf cell. Find the distance ($\delta s_{4}$) between
	the two intersections. 
	Compute \( d\tau _{4}=\chi _{4}\delta s_{4} \). 
	Save  \( d\tau _{4} \).
\item 	There is no intersection, and this is a leaf cell. Do nothing.
\item 	There are two intersections, and this is a leaf cell. 
	Find the distance between ($\delta s_{5}$) the two intersections. 
	Compute \( d\tau _{5}=\chi _{5}\delta s_{5} \).
	Save \( d\tau _{5} \).
	
\end{enumerate}

After the operation described above is finished, the summation in
Eq.~\ref{eq:tau} is performed to find the optical depth between the
current position of the photon and the outer boundary. 

If the cell sizes and their locations are chosen properly, the accuracy of the
optical depth value is within a few percent in normal runs of our models. This
method is used because of the simplicity and the speed. On the other hand, if
accuracy of the optical depth value is crucial, we can use neighboring cell
values to improve the optical depth estimate. Computing the optical depth is
the most time consuming part of the code.

\section{Tests}

\label{sec:Tests}

\subsection{Continuum Polarization Tests}

\begin{figure}
{\par\centering \resizebox*{!}{0.3\textheight}{\includegraphics{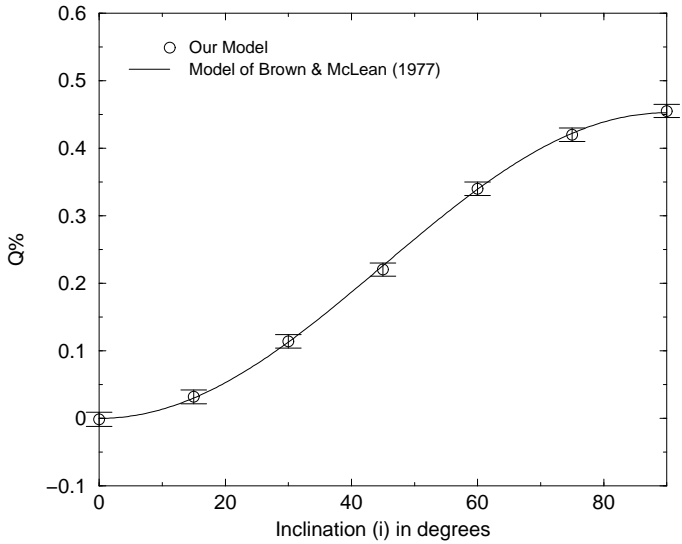}} \par}

\caption{\label{fig:single} Comparison of the polarization v.s.\ inclination angle
of a oblate atmosphere around a single star according to the model of \citet{brown:1977}
(lines) and our model (circles). The model consist of a point source embedded
in the density in the form of Eq.~\ref{eq:denisty}. The result from our model
agrees with the analytic solution of \citet{brown:1977} very well.}
\end{figure}

Firstly, the continuum polarization arising from an optically thin envelope
of a single star is examined. The light source is assumed to be a point source
which is embedded in the density of the following form.

\begin{equation}
\label{eq:denisty}
\rho (r,\theta )=\rho _{o}\, \frac{1}{r^{2}}\, a\, \left( 1+b\cos ^{2}\theta \right) \, 
\end{equation}
 where \( \rho _{o} \) is a constant, \( a \) is an angular normalization
constant and \( b \) is the parameter which controls the angular dependency of
\( \rho  \). When \( b \) is \( <0 \), \( =0 \), and \( >0 \), the stellar
envelope is oblate, spherical, and prolate respectively. The opacity of the
atmosphere is set to be \( \tau \approx 0.1 \). In Fig.~\ref{fig:single},
the normalized polarization levels computed as a function of inclination angle
are shown, and they are compared with the predictions from the model of \citet{brown:1977}.
\( b=-0.5 \) is used in both models. The figure shows that the results from
the two models are in excellent agreement. 

\begin{figure}
{\par\centering \resizebox*{!}{0.3\textheight}{\includegraphics{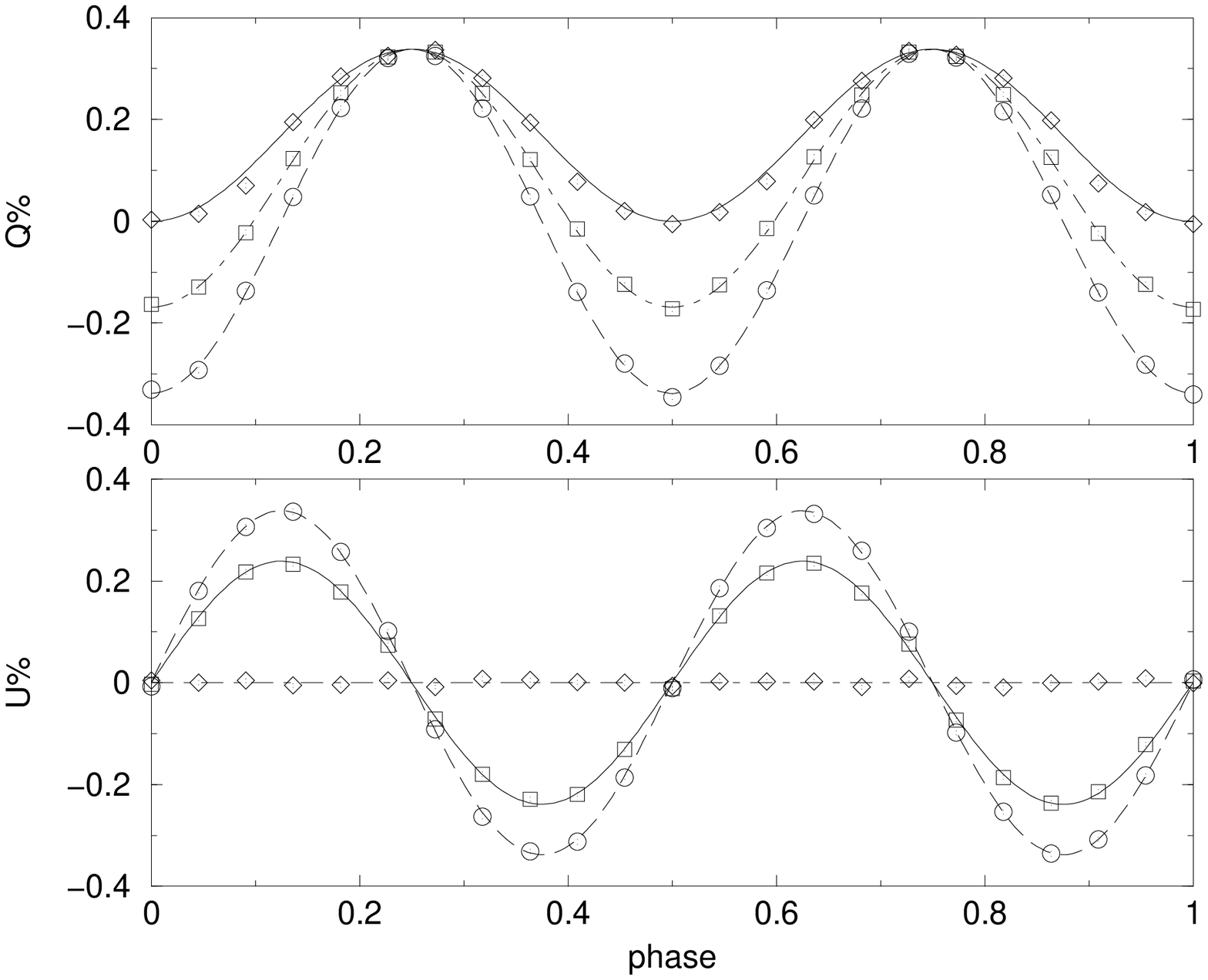}} \par}

\caption{\label{fig:TwoSphere} Comparison of Q \& U polarization double sine curves
predicted by the analytic model of \citet{brown:1978} and our models for three
different inclination angles. Two identical point sources, which are orbiting
around each other in a circular orbit, are each embedded in an optically thin
spherical envelope. The points with makers are from our model, and the lines
are from the model of \citet{brown:1978}: Circles follow \protect\( i=0^{\circ }\protect \)
(face-on) curve, squares follow \protect\( i=45^{\circ }\protect \), and diamonds
follow \protect\( i=90^{\circ }\protect \) (edge-on) curve. The agreement of
the two models is excellent for all inclination angles. The small deviations
of our model from that of \citet{brown:1978} around phase \protect\( =0.42\protect \)
and \protect\( 0.58\protect \) for \protect\( i=90^{\circ }\protect \) is
due to the atmospheric eclipse since the model of \citet{brown:1978} did not
include the attenuation of the star light before scatterings.}
\end{figure}

Secondly, a binary system which consists of two identical point sources orbiting
in a circular orbit is considered. Around each light source, a uniform spherical
density is placed. Again, the envelopes are set to be optically thin so that
the variation of polarization level can be compared with the binary polarization
model of \citet{brown:1978}. Fig.~\ref{fig:TwoSphere} shows that the
results from the two models agree each other very
well for three inclination angles (\( i=0^{\circ },\, 45^{\circ },\, 90^{\circ } \)). 

Thirdly, we verified the computations against the 2-D Monte Carlo polarization
model of \citet{hillier:1991} and the 2-D radiative transfer codes of \citet{hillier:1994}
whose models allow for multiple scatterings (optically thick) and an extended
source. Our results are in good agreement with both models i.e., \( I \) and
\( Q \) are within a fractional error of 0.01.

\subsection{Line Polarization Tests}

\begin{figure}
{\par\centering \resizebox*{!}{0.3\textheight}{\includegraphics{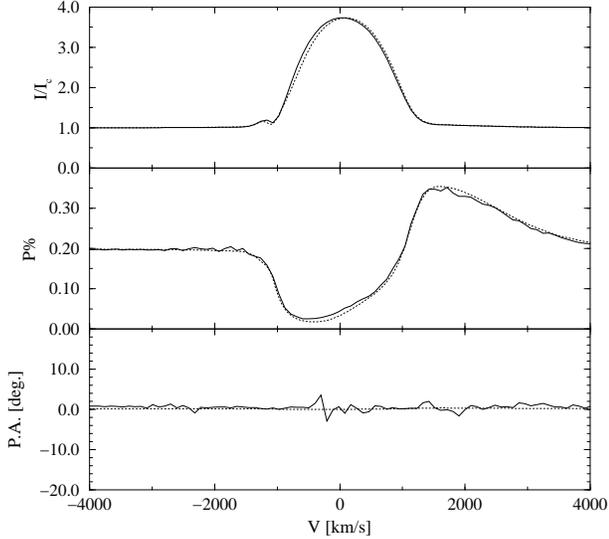}} \par}

\caption{\label{fig:HeIIPol} A model of an optically thick \ion{He}{ii} (\protect\( \lambda 5411\protect \))
line for a WN5 type star. Normalized flux (top), percentage polarization (middle)
and polarization angle (bottom) were computed by our 3-D Monte Carlo model (solid),
and compared with the 2-D radiative transfer model (dotted) of \citet{hillier:1996a}.
The \emph{angular} distribution \emph{}of electron gas is assumed to be the
same as in Eq.~\ref{eq:denisty} with \protect\( b=-0.5\protect \). The agreement
between the two models is excellent.}
\end{figure}

A \ion{He}{i} (\( \lambda 5411 \)) line emitted from a single WN type star
(\( R=1.5R_{\sun } \), \( L=2.0\times 10^{5}\, L_{\sun } \), \( \dot{M}=0.6\times 10^{-5}\, M_{\sun }\, \mathrm{yr}^{-1} \))
is computed with the 3-D Monte Carlo code and the 2-D radiative transfer codes
of \citet{hillier:1996a}. The angular dependency of the atmosphere is assumed
to be the same as in Eq.~\ref{eq:denisty} with \( b=-0.5 \). Fig.~\ref{fig:HeIIPol}
shows the results from both models are in excellent agreement.

\begin{figure}
{\par\centering \resizebox*{!}{0.3\textheight}{\includegraphics{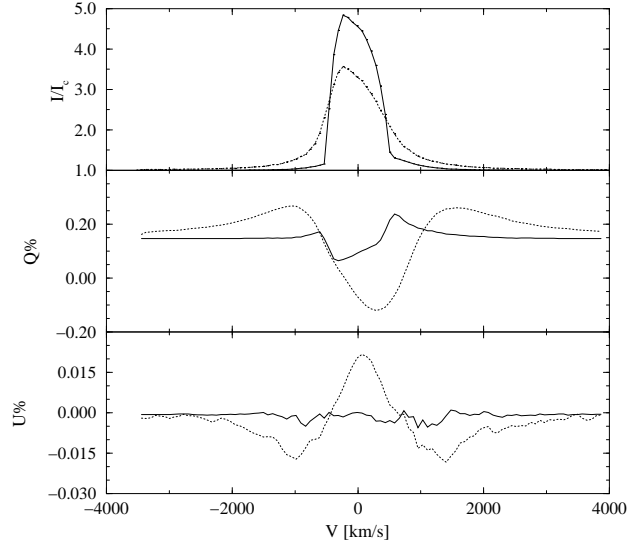}} \par}

\caption{\label{fig:rotation} The effect of rotation on line polarization is demonstrated
in this diagram. Solid line: an emission line produced by a radially expanding
atmosphere. Dotted line: same emission line, but solid-body rotation is added
to the expanding atmosphere. An inclination angle of \protect\( 60^{\circ }\protect \)
was used for both models. The \emph{angular} distribution \emph{}of electrons
was assumed to be the same as in Eq.~\ref{eq:denisty} with \protect\( b=-0.5\protect \). }
\end{figure}

An example of a non-axisymmetric system is a rotating atmosphere of a single
star. Although density distribution is usually symmetric around a rotation axis,
the velocity field is not axisymmetric. To examine a basic behavior of polarization
variations across an emission line, the following two simple models are computed.
1.~an expanding atmosphere (beta velocity law with \( \beta =0.5 \), \( V_{\infty }=500\, \mathrm{km}\, \mathrm{s}^{-1} \))
and 2.~an expanding \& rotating atmosphere (beta-velocity law\( + \)solid-body
rotation with \( V_{\mathrm{rot}}=182\, \mathrm{km}\, \mathrm{s}^{-1} \) on
the stellar surface). The results are shown in Fig.~\ref{fig:rotation}. Qualitatively,
the behavior of these line variations are the same as those described by \citet{mclean:1979c}
(in their Figure~5), and those of a simple analytical model by \citet{wood:1993}
(in their Figs.~4, 6 and 8).

\section{Example Calculations for a Binary System: V444~Cyg}

\label{sec:V444Example}

As an application to a real system, the phase dependent Stokes parameters \( \left( I,\, Q,\, U\right)  \)
of the massive binary system \object{V444~Cyg} (WN5+O6 III-V) are computed
with our model. The system is a short-period (\( P=4.212 \) days, \citealp{khaliullin:1984})
eclipsing binary, and it exhibits variability in polarization, line strength
and X-ray flux as a function of orbital phase. The variability arises from occultation
of the photosphere, from perturbations induced in the extended atmosphere of
the W-R star by the O star and its wind, and from the wind-wind interaction
region. Despite the complexities, many authors \citep[e.g., ][and etc.]{hamann:1992, stlouis:1993, marchenko:1994, cherepashchuk:1995, moffat:1996, marchenko:1997, stevens:1999}
have used this object to determine fundamental parameters of the W-R star by
taking advantage of its variable nature. 

\begin{figure}
{\par\centering \resizebox*{!}{0.3\textheight}{\includegraphics{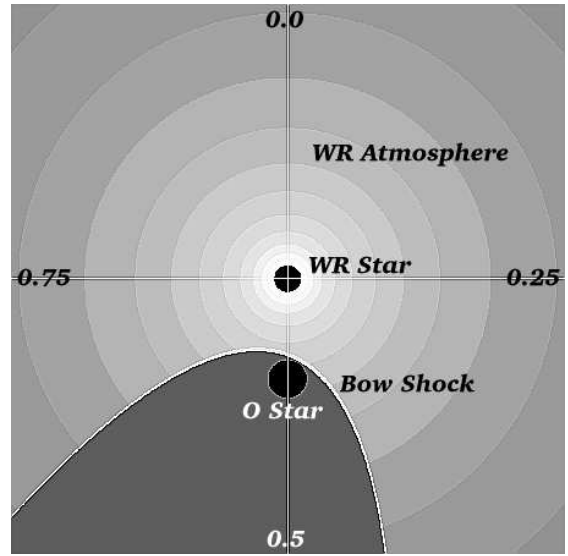}} \par}

\caption{\label{fig:ModelConfig} This figure illustrates the model configuration of
\object{V444~Cyg} (WN5+O6 III-V) on the orbital plane. The W-R star is placed
at the center of the cubic boundary, and it is surrounded by the spherical atmosphere.
The O star is located below the W-R star in the diagram, and the tilted paraboloid
shock with a given thickness is covering the O star. The strong stellar wind
from the W-R star is interrupted by the shock front, and the density behind
the shock is assumed to be insignificantly small for simplicity. The direction
to an observer at phase \protect\( =0,\, 0.25,\, 0.5,\, 0.75\protect \) are
indicated at the top, bottom, left and right edges since the orbital inclination
is about \protect\( 80^{\circ }\protect \) (almost edge-on view). }
\end{figure}

\begin{figure}
{\par\centering \resizebox*{!}{0.3\textheight}{\includegraphics{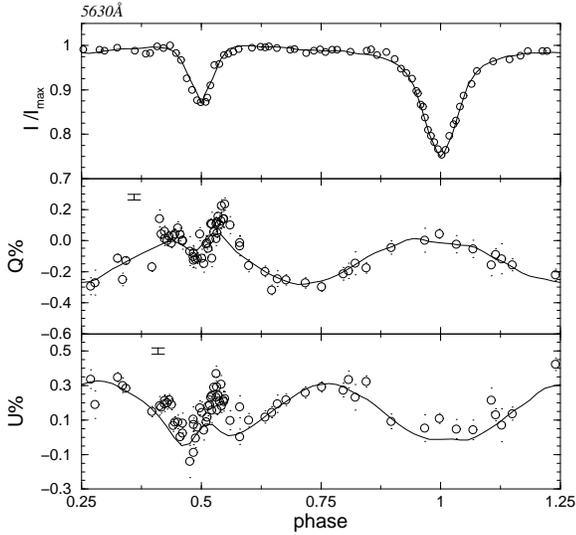}} \par}

\caption{\label{fig:V444PolExample} The figure shows the model calculation of polarization
light curves for V444 Cyg. The relative flux (top), \protect\( Q\protect \)
(middle) and \protect\( U\protect \) (bottom) polarization at \protect\( \lambda =5630\protect \)\AA,
as a function of binary phase, are plotted. Circle: observation, Solid: our
model with \protect\( \dot{M}=0.6\times 10^{-5}\, M_{\sun }yr^{-1}\protect \).
The optical light curve data and optical polarization data are from \citet{kron:1943}
and \citet{stlouis:1993} respectively.}
\end{figure}

\begin{figure}
{\par\centering \resizebox*{!}{0.29\textheight}{\includegraphics{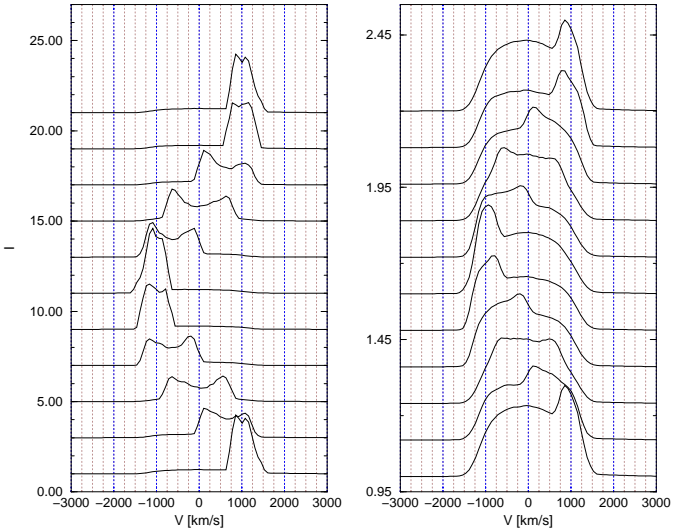}} \par}

\caption{\label{fig:CWemission} Left: Shows the phase dependent emission from the bow
shock region (see Fig.~\ref{fig:ModelConfig}) seen on the top of \ion{He}{i}
(\protect\( \lambda 5876\protect \)) line. The emissivity of the bow shock
region is set to be unrealistically strong in order to demonstrate the effect
of the bow shock emission. Right: Same as for the left figure, but the emissivity
of the bow shock region is reduced by a factor of \protect\( \sim 10\protect \).
This figure qualitatively describes the variability seen in the observation
of this line \citep[see e.g., ][]{marchenko:1994}. The profiles are plotted
from phase \protect\( =0\protect \) (bottom) to phase \protect\( =1\protect \)
(top) with \protect\( 0.1\protect \) phase steps.}
\end{figure}

The model consists of three main components: 1.~the W-R star
atmosphere, 2.~the O~star's spherical surface 
(with a limb-darkening law) and 3.~the paraboloid-shaped
bow shock region due to the colliding stellar winds (Fig.~\ref{fig:ModelConfig}).
More detailed discussion on the \object{V444~Cyg} polarization model can be
found in \citet{kurosawa:2001b}. Fig.~\ref{fig:V444PolExample} shows the resulting
continuum Stokes parameters \( \left( I,\, Q,\, U\right)  \) at \( 5630 \)\AA~as
a function of orbital phase. The model reproduces the eclipsing \( I \)
light curve, the double sine curves of \( Q \) \& \( U \) polarizations and the
small oscillations seen in \( Q \) \& \( U \) near the secondary eclipse (phase=0.5).
Next, the phase dependent \ion{He}{i} (\( \lambda  \)5876) emission line, created
mainly in the outer part of the W-R atmosphere in the binary, is computed. The
results are shown in Fig.~\ref{fig:CWemission}. The left plot in Fig.~\ref{fig:CWemission}
shows the sequence of \ion{He}{i} (\( \lambda  \)5876) line profiles computed
with unrealistically high emissivity in the bow shock region, to emphasize the
effect of the bow shock. The underlying ordinary \ion{He}{i} emission from the
W-R star is barely visible in this plot. On the left in Fig.~\ref{fig:CWemission},
the sequence of the same line is displayed, but with a more realistic amount
of emission from the bow shock region. This figure qualitatively describes the
variability seen in the observation of this line \citep[see e.g., ][]{marchenko:1994}.

\section{Conclusions}

\label{sec:Conclusion}

We have presented a new 3-D Monte Carlo model which computes the variability
in line and continuum polarization associated with the orbital motion of a binary
system surrounded by a non-axisymmetric envelope. The basic model is constructed
in a general manner so that it can handle an arbitrary distribution of gas density.
A special method of grid spacing called the {}``cell-splitting{}'' method is
used to automatically assign more grid points to the higher density regions.
For a fixed number of grid points, the tree method achieves a higher
accuracy than the uniform cubic grid method.
This kind of grid scheme is very useful in multi-dimensional calculations. It
is not restricted to a binary system, but also applicable to many other astrophysical
systems, e.g., accretion flows, molecular clouds and a 3-D hydrodynamical model.

The model was tested extensively. We demonstrated the accuracy of our model
by comparing it with simple analytical models and existing well-tested 2-D numerical
models. Firstly, the model was tested against the analytic models of continuum
polarization by \citet{brown:1977} and \citet{brown:1978}, and they were in
excellent agreement for both a point source and a binary model. Secondly, the
polarization in the \emph{optically thick} \ion{He}{ii} (\( \lambda 5411 \))
line calculated by our model and by the 2-D radiative transfer model of \citet{hillier:1996a}
were compared. They also showed good agreement. Thirdly, the effect of rotation
on the line polarization seen in an emission line was demonstrated. The basic
behavior of polarization and polarization angle across an emission line, computed
by our model, are confirmed to be the same as those in \citet{mclean:1979c}
and \citet{wood:1993}. 

We also demonstrated the application of our model to a real system: \object{V444~Cyg}.
The model was able to produce a set of complicated \( I \), \( Q \) and \( U \)
light curves which fit the observational data very well. In addition, we qualitatively
modeled the phase dependent excessive emission seen on the top of the optically
thin \ion{He}{i} (\( \lambda 5976 \)) line. Since this excessive emission is
originated from the bow shock heated region due to the colliding stellar winds,
by modeling the sequence of this line properly, we would be able to probe little
known properties (e.g., gas flow, geometrical configuration) of the bow shock
in the system.  If one wishes to model a line variability associated with the
bow-shock region correctly, one needs to have realistic opacity information
in the region. A more formal approach to this kind of problem is to develop
a full 3-D non-LTE radiative transfer model. Because of the high dimensionality,
computational time would increase much more than in the case of a 1-D problem.
Developing a 3-D Monte Carlo \emph{non-LTE} radiative transfer model would be
simpler than developing the 3-D short-characteristic ray method e.g., by \citet{folini:1999},
and the tree data structure is relevant to this problem.

In a following paper \citep{kurosawa:2001b}, we will apply our model, in conjunction
with the multi-line non-LTE radiative transfer model of \citet{hillier:1998},
to estimate the mass-loss rate of \object{V444~Cyg}. This will be done by fitting
the observed \ion{He}{i} (\( \lambda 5876 \)) and \ion{He}{ii} (\( \lambda 5412 \))
line profiles, and the continuum light curves of three Stokes parameters \( \left( I,\, Q,\, U\right)  \)
in \( V \) band simultaneously. 

%

%




%

%

%

%

\begin{thebibliography}{40}
\expandafter\ifx\csname natexlab\endcsname\relax\def\natexlab#1{#1}\fi

\bibitem[{Bernes(1979)}]{bernes:1979}
Bernes, C. 1979, \aap, 73, 67

\bibitem[{{Brown} {et~al.}(1989){Brown}, {Carlaw}, \&
  {Cassinelli}}]{brown:1989}
{Brown}, J.~C., {Carlaw}, V.~A., \& {Cassinelli}, J.~P. 1989, \apj, 344, 341

\bibitem[{Brown \& McLean(1977)}]{brown:1977}
Brown, J.~C. \& McLean, I.~S. 1977, \aap, 57, 141

\bibitem[{Brown {et~al.}(1978)Brown, McLean, \& Emslie}]{brown:1978}
Brown, J.~C., McLean, I.~S., \& Emslie, A.~G. 1978, \aap, 68, 415

\bibitem[{{Chandrasekhar}(1960)}]{chandrasekhar:1960}
{Chandrasekhar}, S. 1960, Radiative transfer (New York: Dover, 1960)

\bibitem[{Cherepashchuk {et~al.}(1995)Cherepashchuk, Koenigsberger, Marchenko,
  \& Moffat}]{cherepashchuk:1995}
Cherepashchuk, A.~M., Koenigsberger, G., Marchenko, S.~V., \& Moffat, A.~F.~J.
  1995, \aap, 293, 142

\bibitem[{{Folini} \& {Walder}(1999)}]{folini:1999}
{Folini}, D. \& {Walder}, E. 1999, in IAU Symp., Vol. 193, Wolf-Rayet Phenomena
  in Massive Stars and Starburst Galaxies, ed. K.~A. van~der Hucht,
  G.~Koenigsberger, \& P.~R.~J. Eenens, 352

\bibitem[{{Gayley} {et~al.}(1997){Gayley}, {Owocki}, \&
  {Cranmer}}]{gayley:1997}
{Gayley}, K.~G., {Owocki}, S.~P., \& {Cranmer}, S.~R. 1997, \apj, 475, 786

\bibitem[{{Hamann} \& {Schwarz}(1992)}]{hamann:1992}
{Hamann}, W.\, R. \& {Schwarz}, E. 1992, \aap, 261, 523

\bibitem[{{Harries}(2000)}]{harries:2000}
{Harries}, T.~J. 2000, \mnras, 315, 722

\bibitem[{Hillier(1991)}]{hillier:1991}
Hillier, D.~J. 1991, \aap, 247, 455

\bibitem[{Hillier(1994)}]{hillier:1994}
---. 1994, \aap, 289, 492

\bibitem[{Hillier(1996)}]{hillier:1996a}
---. 1996, \aap, 308, 521

\bibitem[{Hillier \& Miller(1998)}]{hillier:1998}
Hillier, D.~J. \& Miller, D.~L. 1998, \apj, 496, 407

\bibitem[{Juvela(1997)}]{juvela:1997}
Juvela, M. 1997, \aap, 322, 943

\bibitem[{Khaliullin {et~al.}(1984)Khaliullin, Khaliullina, \&
  Cherepashchuk}]{khaliullin:1984}
Khaliullin, K.~F., Khaliullina, A.~I., \& Cherepashchuk, A.~M. 1984, Soviet
  Astron.~Lett., 10, 250

\bibitem[{Koenigsberger \& Auer(1985)}]{koenigsberger:1985}
Koenigsberger, G. \& Auer, L.~H. 1985, \apj, 297, 255

\bibitem[{{Kron} \& {Gordon}(1943)}]{kron:1943}
{Kron}, G.~E. \& {Gordon}, K.~C. 1943, \apj, 97, 311

\bibitem[{Kurosawa \& Hillier(2001)}]{kurosawa:2001b}
Kurosawa, R. \& Hillier, D.~J. 2001, \apj, submitted

\bibitem[{{Li} \& {McCray}(1996)}]{li:1996}
{Li}, H. \& {McCray}, R. 1996, \apj, 456, 370

\bibitem[{Lucy(1999)}]{lucy:1999}
Lucy, L.~B. 1999, \aap, 345, 211

\bibitem[{Luo {et~al.}(1990)Luo, McCray, \& Mac~Low}]{luo:1990}
Luo, D., McCray, R., \& Mac~Low, M. 1990, \apj, 362, 267

\bibitem[{Marchenko {et~al.}(1997)Marchenko, Moffat, Eenens, Cardona,
  Echevarria, \& Hervieux}]{marchenko:1997}
Marchenko, S.~V., Moffat, A.~F.~J., Eenens, P.~R.~J., Cardona, O., Echevarria,
  J., \& Hervieux, Y. 1997, \apj, 422, 810

\bibitem[{Marchenko {et~al.}(1994)Marchenko, Moffat, \&
  Koenigsberger}]{marchenko:1994}
Marchenko, S.~V., Moffat, A.~F.~J., \& Koenigsberger, G. 1994, \apj, 422, 810

\bibitem[{McLean(1979)}]{mclean:1979c}
McLean, I.~S. 1979, \mnras, 186, 265

\bibitem[{{Modali} {et~al.}(1972){Modali}, {Brandt}, \&
  {Kastner}}]{modali:1972}
{Modali}, S.~B., {Brandt}, J.~C., \& {Kastner}, S.~O. 1972, \apj, 175, 265

\bibitem[{{Moffat} \& {Marchenko}(1996)}]{moffat:1996}
{Moffat}, A. F.~J. \& {Marchenko}, S.~V. 1996, \aap, 305, L29

\bibitem[{Pagani(1998)}]{pagani:1998}
Pagani, L. 1998, \aap, 333, 269

\bibitem[{Pittard(1998)}]{pittard:1998}
Pittard, J.~M. 1998, \mnras, 300, 479

\bibitem[{Pittard \& Stevens(1999)}]{pittard:1999}
Pittard, J.~M. \& Stevens, I.~R. 1999, in IAU Symp., Vol. 193, Wolf-Rayet
  Phenomena in Massive Stars and Starburst Galaxies, ed. K.~A. van~der Hucht,
  G.~Koenigsberger, \& P.~R.~J. Eenens (San Francisco, Calif: Astronomical
  Society of the Pacific), 386

\bibitem[{{Sandford}(1973)}]{sandford:1973}
{Sandford}, M.~T. 1973, \apj, 183, 555

\bibitem[{Shore \& Brown(1988)}]{shore:1988}
Shore, S.~N. \& Brown, D.~N. 1988, \apj, 334, 1021

\bibitem[{St-Louis {et~al.}(1993)St-Louis, Moffat, Lapointe, Efimov,
  Shakhovskoj, Fox, \& Piirola}]{stlouis:1993}
St-Louis, N., Moffat, A.~F.~J., Lapointe, L., Efimov, Y.~S., Shakhovskoj,
  N.~M., Fox, G.~K., \& Piirola, V. 1993, \apj, 410, 342

\bibitem[{Stevens {et~al.}(1992)Stevens, Blondin, \& Pollock}]{stevens:1992}
Stevens, I.~R., Blondin, J.~M., \& Pollock, A.~M.~T. 1992, \apj, 386, 265

\bibitem[{{Stevens} \& {Howarth}(1999)}]{stevens:1999}
{Stevens}, I.~R. \& {Howarth}, I.~D. 1999, \mnras, 302, 549

\bibitem[{Stevens \& Pollock(1994)}]{stevens:1994}
Stevens, I.~R. \& Pollock, A.~M.~T. 1994, \mnras, 269, 226

\bibitem[{{Warren-Smith}(1983)}]{warrensmith:1983}
{Warren-Smith}, R.~F. 1983, \mnras, 205, 337

\bibitem[{Witt \& Gordon(1996)}]{witt:1996}
Witt, A.~N. \& Gordon, K.~D. 1996, \apj, 463, 681

\bibitem[{Wolf {et~al.}(1999)Wolf, Henning, \& Stecklum}]{wolf:1999}
Wolf, S., Henning, T., \& Stecklum, B. 1999, \aap, 349, 839

\bibitem[{{Wood} {et~al.}(1993){Wood}, {Brown}, \& {Fox}}]{wood:1993}
{Wood}, K., {Brown}, J.~C., \& {Fox}, G.~K. 1993, \aap, 271, 492

\end{thebibliography}

%

%

%
\end{document}